\documentclass[useAMS,usenatbib]{mn2e}

\usepackage{amsmath,amssymb,amsfonts,latexsym}
\usepackage[dvips]{graphicx}
\usepackage{longtable}
\usepackage{fixltx2e}
\usepackage{lscape}
\title[Pulse-width statistics in radio pulsars]
{On the pulse-width statistics in radio pulsars.\\ III. Importance of the conal profile components.}

\author[Krzysztof Maciesiak, Janusz Gil \& Giorgi Melikidze]{Krzysztof Maciesiak$^{1}$\thanks{E-mail:jezyk@astro.ia.uz.zgora.pl}, Janusz Gil$^{1}$, Giorgi Melikidze$^{1,2}$\\
$^{1}$Kepler Institute of Astronomy, University of Zielona G\'{o}ra, Lubuska 2, 65-265 Zielona G\'{o}ra, Poland \\
$^2$ Ilia State University, E. Kharadze Abastumani Astrophysical Observatory, Tbilisi, Georgia}
\begin{document}

\date{Accepted . Received ; in original form }

\pagerange{\pageref{firstpage}--\pageref{lastpage}} \pubyear{2011}

\maketitle

\label{firstpage}

\begin{abstract}
This work is a continuation of two previous papers of a series, in which we examined the pulse-width statistics of normal radio pulsars. In the first paper we compiled the largest ever database of pulsars with interpulses in their mean profiles. In the second one we confirmed the existence of the lower boundary in the scatter plot of core component pulse-widths versus pulsar period $W_{50}\sim2.5^{\circ}P^{-0.5}$, first discovered by Rankin using much smaller number of interpulse cases. In this paper we show that the same lower boundary also exists for conal profile components. Rankin proposed a very simple method of estimation of pulsar inclination angle based on comparing the width $W_{50}$ of its core component with the period dependent value of the lower boundary. We claim that this method can be extended to conal components as well. To explain an existence of the lower boundary Rankin proposed that the core emission originates at or near the polar cap surface. We demonstrated clearly that no coherent pulsar radio emission can originate at altitudes lower than 10 stellar radii, irrespective of the actual mechanism of coherence. We argue that the lower boundary reflects the narrowest angular structures that can be distinguished in the average pulsar beam. These structures represent the core and the conal components in mean pulsar profiles. The $P^{-0.5}$ dependence follows from the dipolar nature of magnetic field lines in the radio emission region, while the numerical factor of about $2.5^{\circ}$ reflects the curvature radius of a non-dipolar surface magnetic field in the partially screened gap above the polar cap, where dense electron-positron plasma is created. Both core and conal emission should originate at altitudes of about 50 stellar radii in a typical pulsar, with a possibility that the core beam is emitted at a slightly lower heights than the conal ones.
\end{abstract}


\section{Introduction}
\begin{figure}
\begin{center}
\includegraphics{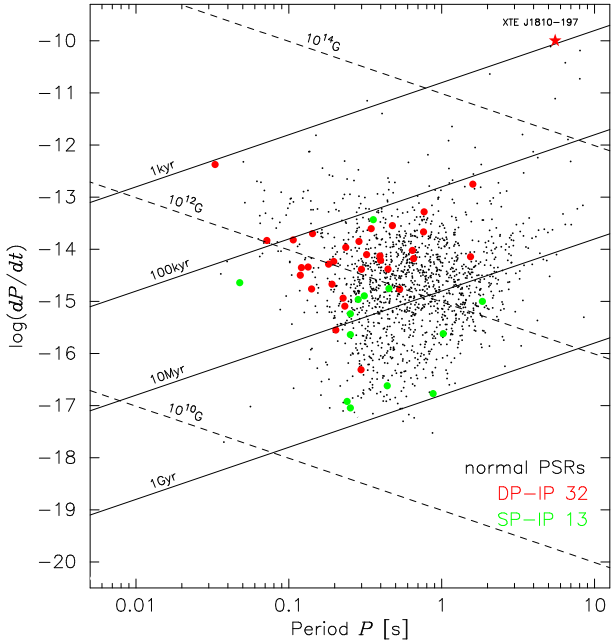}
\caption{Diagram $P-\dot{P}$ for normal radio pulsars, including 45 cases with interpulse emission. Black dots represent 1450 normal pulsars, while red and green circles correspond to 32 double-pole (DP-IP) and 13 single-pole (SP-IP) interpulsars, respectively. Lines of constant magnetic field and constant characteristic age are shown. Notice the special case marked by red star representing the radio emitting magnetar XTE J1810-197, which shows narrow DP-IP feature (see also its location in Figure \ref{figure.2}). Notice that DP-IP (almost orthogonal) and SP-IP (almost aligned) cases are clearly separated (see also) Figure \ref{figure.2}. \label{figure.1}}
\end{center}
\end{figure}

In recent years there were performed a number of major campaigns to search pulsars. The most important surveys that found the largest numbers of new pulsars were Parkes Multibeam Pulsar Surveys (\citet{manchester01,morris02,kramer03,hobbs04,faulkner04,lorimer06}; PMPS I -- VI hereafter). The important basic parameters of these new pulsars are easily available either from the discovery papers or from the ATNF Pulsar Catalogue\footnote{http://www.atnf.csiro.au/research/pulsar/psrcat/} \citep{manchester05}. Maciesiak, Gil \& Ribeiro (2011; hereafter \citet{mgr11}) and Maciesiak \& Gil (2011; hereafter \citet{mg11}) explored different aspects of the pulse-width statistics in normal (non-millisecond and other non-recycled) pulsars. The most important result of \citet{mgr11} was the compilation of the largest ever database of 44 pulsars with interpulse emission, including 31 double-pole (DP-IP) and 13 single-pole (SP-IP) interpulsars, respectively. These pulsars when analysed on the $P-\dot{P}$ diagram revealed a clear evolutionary tendency. Namely, SP-IP cases were on average about 3 times older than DP-IP cases. In addition, SP-IP cases were also representing on average about 3 times weaker magnetic fields than DP-IP cases. This is illustrated in Figure \ref{figure.1}, which reproduces the $P-\dot{P}$ diagram from \citet{mgr11} with the important addition of radio magnetar XTE J1810-197. This object shows many typical properties of radio pulsars, including the emission of DP-IP. It is reasonable to assume that DP-IP and SP-IP cases represent almost orthogonal and almost aligned rotators, respectively. Based on this assumption it was concluded in \citet{mgr11} that the apparent separation of DP-IP and SP-IP cases on the $P-\dot{P}$ diagram revealed a secular alignment of the magnetic axis towards the spin axis with a random initial value of the inclination angle. The radio emitting magnetar XTE J1810-197 fits perfectly to this scheme.

\begin{figure}
\begin{center}
\includegraphics{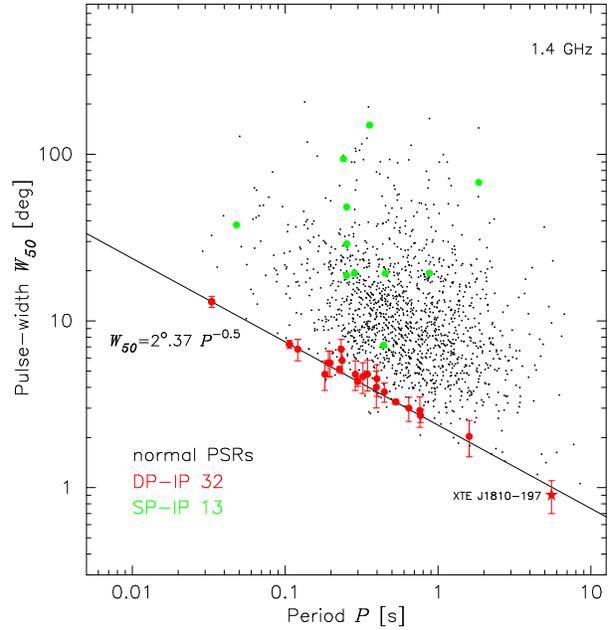}
\caption{Plot of the 1.4 GHz pulse-width $W_{50}$ versus the period $P$ for 1450 normal pulsars from the ATNF database. Superposed are 32 DP-IP cases (marked in red) and 13 SP-IP cases (marked in green) taken from Table 1 and Figure 2 in \citet{mg11} and rescaled to 1.4 GHz. Notice the special DP-IP case marked by red star, which represents the width of a narrow component in the radio profile of the magnetar XTE J1810-197. Notice that DP-IP (almost orthogonal) and SP-IP (almost aligned) cases are clearly separated (see also) Figure \ref{figure.1}.
\label{figure.2}}
\end{center}
\end{figure}

\citet{mgr11} emphasized importance of the interpulses in the pulse-width statistics, while in \citet{mg11} importance of the core profile components was demonstrated. In particular, the relationship $W_{50}=2.45^{\circ} P^{-0.5}$ for 50 per cent maximum intensity pulse-widths of core profile components found by Rankin (1990; \citet{r90} hereafter) in 6 interpulsars was confirmed using 21 pulsars with double-pole interpulses for which measurement of the core component width was possible. \citet{r90} claimed that this relationship represents the Lower Boundary Line (LBL hereafter) in the scatter plot of $W_{50}$ versus $P$, and we confirmed this in \citet{mg11} (see red marked symbols in Figure \ref{figure.2}). 

In this paper we argue that the LBL relies upon pulse-widths of both core and conal components of pulsar profiles. However, one should realize that the lower boundary is not as sharp as the formal fitting to the red marked interpulse data points in Figure \ref{figure.2} suggests. It can be expressed approximately in the form $W_{50}= (2.4^{\circ}\pm 1^{\circ})P^{-0.5}$ (see discussion of LBB below Figure 8 in \citet{mg11}). Based on the existence of the LBL \citet{r90} proposed a simple method of estimation of inclination angle $\alpha=\arcsin(2.45^{\circ} P^{-0.5}/W_{50})$ in pulsars with core component. In this paper we argue that this method can be applied to the conal components as well.

The case of interpulse in radio magnetar XTE J1810-197 deserves a special attention. A sporadic pulsed radio emission of this magnetar is present only during the outburst phase \citep{kramer07}. The presence of a narrow interpulse is consistent with the assumption of an almost orthogonal rotator. Consequently, its location in Figure \ref{figure.1} (marked by the red star) characteristic of DP-IP cases is well understood. Because of the time varying profile shape it is difficult to measure pulse-widths of different components. Nevertheless, it seems that the narrowest feature in the mean profile has the half-power width $W_{50}=0.9^{\circ}\pm0.2^{\circ}$ (Figure 3 in \citet{kramer07} and Figure 1 in \citet{serylak09}). We marked this measurement by the red star in Figure \ref{figure.2}. It is very interesting that even such exotic object as radio emitting magnetar with interpulse follows the LBL characteristic for core components of DP-IP pulsars.

\begin{figure*}
\includegraphics{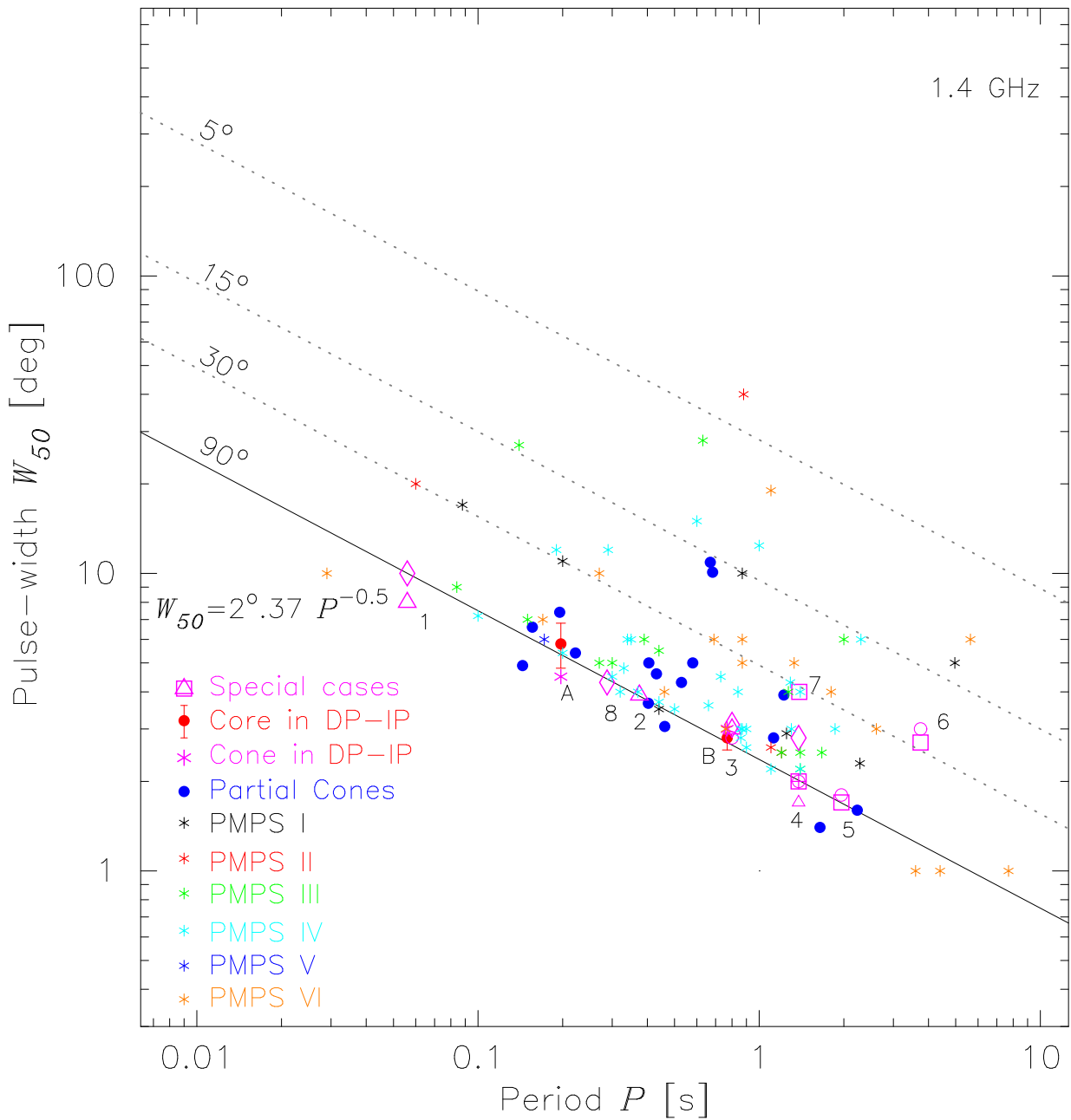}
\vbox to 20mm{\vfil\caption{Scatter plot of the 50 per cent maximum intensity pulse-width $W_{50}$ of conal profile components versus the pulsar period $P$ for 96 cases, in which the appropriate estimates from 1.4 GHz profiles were possible (Table \ref{table.1}). Different colours represent different blocks of data marked in the figure key and specified in the text. In two DP-IP cases marked by A and B symbols both core and conal component widths are presented. The dotted lines representing different inclination angles $\alpha$ were computed using the method of \citet{r90}.
\label{figure.3}} \vfil}
\end{figure*}

The clear evidence of the LBL in the 50 per cent pulse-width $W_{50}=2.45^{\circ}\nu_\text{GHz}^{-0.1} P^{-0.5}$ is very intriguing and requires a solid explanation. \citet{r90} argued that the core component widths are intimately related to the polar cap geometry at the stellar surface, which under some assumptions has a similar angular opening $2\rho=2.49^{\circ} P^{-0.5}$, which can be just a pure coincidence \footnote{Indeed, one should realise that $2\rho=2.49^{\circ}P^{-0.5}$ holds only for the \citet{goldreich69} polar cap with radius $r_p=1.4\times10^4 P^{-0.5}$ cm, where $\rho$ is the angle between the pulsar dipolar axis and tangent to the last open dipolar magnetic field line at the polar cap surface. In reality this is rather poor description of the locus of the last open field lines at the polar cap. Instead of the light cylinder radius $R_{LC}=cP/2\pi$ defining the last open dipolar field lines, one should rather consider the corotation limiting radius, which can be significantly smaller than $R_{LC}$. Consequently, the polar cap radius can be significantly different from the value of $r_p$. Also, in the realistic force-free magnetospheric models (e.g. \citet{spitkovsky06}) the size of the polar cap differs from $r_p$.}. Although this natural and very appealing explanation is commonly accepted, we have a number of doubts and reservations about it, which are reviewed and discussed critically in Section \ref{subsection.r90review}. One minor problem is the apparent frequency dependence of $W_{50}$ of core components, at least in DP-IP cases\footnote{The fact that the LBLs of $W$ are consistent with $\nu_\text{GHz}^{-0.1}$ frequency dependence is interesting and worth of short discussion. In general $W_{50} \propto \rho \propto r^{0.5} \propto \nu^{-0.1}$ (see Equations (2) and (3) in \citet{kg98}). Although, we used these equations to rescale the numerical factor in the LBL as well as the interpulse data between 1.0 and 1.4 GHz, it seems that the LBL follows the frequency dependence similar to that of the pulse profile widths. This is disfavouring the \citet{r90} interpretation of the core emission originating at the polar cap surface, independent of the radiation frequency.}. The more serious problem is presented in Section \ref{section.constraints}, where we show using a basic physical arguments that no coherent radio emission (no matter core or conal type) can originate at or near the polar cap surface (at altitudes lower than about 10 neutron star radii). Thus, the core component cannot be emitted at or near the polar cap, and new explanation and interpretation of the lower boundary in pulse-width data is required. In this paper we attempt to provide an alternative explanation and discuss its consistence with both the observational data and physical models of the inner acceleration gap in pulsars.

\section{Importance of the conal profile components}\label{section.2}

%\begin{figure}
%\begin{center}
%\includegraphics{fig4.eps}
%\caption{\textbf{Plot of $\Delta W = W_{10} - W_{50}$ versus pulsar period $P$, where $W_{10}$ and $W_{50}$ are pulse-widths measured at 10 and 50 per cent of maximum intensity, respectively (data for 908 normal pulsars from ATNF database with both $W_{10}$ and $W_{50}$ measurements). The lower boundary $1.4^{\circ} P^{-0.5}$ was derived from the analysis of conal component widths (e.g. \citet{gks93}). The dashed line representing the LBL was added for the reference.
%\label{figure.4}}}
%\end{center}
%\end{figure}
%figure new 4

Figure \ref{figure.2} contains 1450 measurements of $W_{50}$ of normal pulsars imported blindly from the ATNF database, without any distinction into core or conal components. Apparently, the bulk of the data points corresponds to the full profile widths. For the clarity of considerations we superposed on this figure the appropriate measurements of interpulse cases (colour filled circles). The almost orthogonal rotators represented by DP-IP cases marked in red lie near the LBL $W_{50}=2.37^{\circ} P^{-0.5}$. It is important to know whether the other data points (black dots) near the LBL also represent the widths of the core components in nearly orthogonal rotators with IP emission missing for some reasons. As one can judge from Figure 5 in \citet{mg11} the vicinity of the LBL (called the Lower Boundary Belt) includes pulsars with $45^{\circ}<\alpha<135^{\circ}$. It is then easy to understand why interpulses are not visible for most of the points near the LBL. Indeed, as given by the detection condition for DP-IP cases $\rho>180^{\circ}-\beta-2\alpha$ (for details see Equation (21) in \citet{mgr11}) $\alpha$ must be really close to $90^{\circ}$ (within several degrees) for the interpulse to be seen.

We know from \citet{mg11} that some partial cone profiles (in which $W_{50}$ is determined by one dominating conal component) lie in the vicinity of the LBL (see Figure 3 there). Thus, perhaps the conal component widths also scale like that of the core components, that is $W_{50}\sim2.5^{\circ} P^{-0.5}/\sin\alpha$ (e.g. \citet{r90}). To resolve this problem we present in Figure \ref{figure.3} the scatter plot of $W_{50}$ versus $P$, using only measurements of a strong conal components. In the Appendix \ref{appendix.spec} we discuss different blocks of data listed in Table \ref{table.1} and presented in Figure \ref{figure.3}. The important comment that should be made here is that these data represent all cases in which we were able to make a credible $W_{50}$ measurement/estimate of the conal component(s) from the profiles available in the literature (marked in column Ref. of Table \ref{table.1}).

As one can see from Figure \ref{figure.2} the distribution of the conal width data is pretty much the same as that of the core width data presented by \citet{r90} (see Figures 1 and 3 there) and confirmed in our \citet{mg11} (see Figure 2 there). This is also a reason why the mixed distribution presented in Figure \ref{figure.2} looks similar too, although it contains measurements of both core and conal components, as well as full profile widths. The most characteristic feature of these distributions is a clear presence of the LBL with the slope proportional to $P^{-0.5}$. The proportionality factor about $2.5^{\circ}$ is apparently the same for the core and the conal components. The $P^{-0.5}$ slope can be understood in terms of divergence of dipolar magnetic field lines in the radio emission region, while the actual value of the proportionality factor is not quite clear. It will be argued in this paper that the occurrence of the LBL in pulse-width distribution represents the narrowest angular structures that can be distinguished in the average pulsar beam. These structures represent components in the mean pulsar profile (no matter core or conal type), whose associated emission counterparts fill about the same angular volume of the overall mean pulsar beam, independently of the pulsar period. The pulsar radio emission is a complicated multi-step process that begins at the polar cap in the form of sparks discharging the ultra-high potential drop of the inner pulsar accelerator. As argued in Section \ref{section.beam.structures} each spark is a very narrow entity with a size of just few meters across it. The plasma produced by each of these sparks follows the diverging magnetic field lines. We estimated that angular extent of each spark-associated plasma flux tube is about $2.5^{\circ}P^{-0.5}$ when it reaches the radio emission region. We claim that this explains the existence of the LBL.

\section{Explanation of the LBL}\label{section.explanation.r90}

\subsection{Critical review of canonical interpretation of the LBL}\label{subsection.r90review}
As already mentioned \citet{r90} proposed a very simple and natural interpretation of the LBL appearing in the pulsar data as $W_{50}\approx2.5^{\circ}P^{-0.5}$. Since the nominal value of the LBL is almost equal to the opening angle of the polar cap $2\rho=2.49^{\circ} P^{-0.5}$ (see footnote 2), Rankin concluded that the core emission must be emitted from the entire surface of the polar cap. More generally one should write $2\rho=2.49^{\circ} s_{50} r_6^{0.5} P^{-0.5}$, where $s_{50}$ is the parameter describing the locus of field lines corresponding to 50 per cent width of the core component and $r_6$ is the normalised altitude of the core emission in units of star radius $R=10^6$ cm. Apparently, Rankin assumed that both $s_{50}$ and $r_6$ are equal to 1, while the more general conclusion would be $s_{50} r_6^{0.5}\approx1$ (e.g. \citet{gil91}). It is assumed that the pulsar radio emission is relativistically beamed tangentially to the open dipolar magnetic field lines into a very narrow solid angle.

Let us discuss the first assumption $s_{50}=1$, which means that 50 per cent maximum intensity of the core component originates at the last open field lines of the polar cap. Then the question arises where does the rest of the core component originate? One can argue that the emission region corresponding to the 50 per cent maximum intensity as well as edge of the profile involve approximately the same dipolar field lines i.e. $s_{50}\approx s_{10}\approx 1$. This argument can work in pulsars with steep core components but not in the case of long tails in core single profiles, for which there must be $s_{50} < s_{10} \sim 1$. Unless this emission is beamed tangentially to the closed field lines, a possibility which does not seem likely.

One can always say that the emission region is slightly above the polar cap, but in some cases the emission altitude would have to be quite large to fit the long tail of the core component (below 50 per cent of the maximum intensity). Thus, it seems that the parameter $s_{50}$ must be considerably smaller than 1.

The other assumption of \citet{r90} that $r_6\approx1$ for core components is also very difficult to justify. First of all, there is an observational evidence of the radius-to-frequency mapping in core component widths (see \citet{mitra04} and footnote 3), which is difficult to explain with the emission originating at the polar cap. Moreover, in the Section \ref{section.constraints} we present the physical arguments showing that no radio emission can originate at low altitudes $r_{em}<10R$. This is the most serious problem for the canonical interpretation, meaning that the core emission cannot originate even close to the polar cap surface. Moreover, with $r_6 > 10$ we get $2\rho > 2.49^{\circ} r_6^{0.5} P^{-0.5} = 7.87^{\circ}P^{-0.5}$ for $s_{50}=1$, which is high above the LBL value $W_{50}=2.45^{\circ}\:P^{-0.5}$. This confirms that indeed $s_{50}$ must be considerably smaller than 1.

\subsection{The new interpretation of the LBL}\label{section.new.interpretation}
\begin{figure}
\begin{center}
\includegraphics{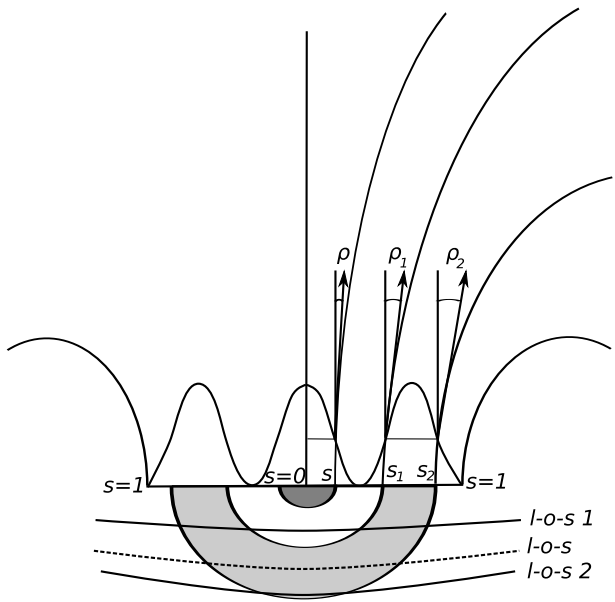}
\caption{Schematic representation of the cross-section of open dipolar field lines at altitudes of the radio emitting region. The parameter $s$ labels different field lines from $s=0$ for dipolar axis to $s=1$ for the last open field lines. Three Gaussian profiles represent three subbeams of instantaneous radio emission associated with three different plasma flows. For simplicity only one conal ring around the core pencil beam is marked. Also, the shapes and mutual separation of subpulses are idealized. The opening angles of emission $\rho$, $\rho_1$ and $\rho_2$ between the magnetic axis and the tangent to dipolar field lines are marked for the level corresponding to $W_{50}$ for each component. The ring of mean conal emission is marked schematically in light grey, while the pencil beam of mean core emission is marked schematically in darker grey (both cross-sections at 50 per cent level). This conal ring is formed by subbeams of instantaneous emission and their carousel motion around the magnetic axis. Two exemplary line of sights (l-o-s 1 and l-o-s 2) giving approximately the same pulse-widths of the observed components are marked. The third one (l-o-s) illustrates the case when the pulse-width can be broader. For very small impact angles one can observe a core component flanked by two conal components, all three of about the same width.
\label{figure.4}}
\end{center}
\end{figure}

Given the critical arguments presented in the previous section, the existence of the LBL in the pulse-width data requires a new interpretation. The $P^{-0.5}$ dependence is obviously associated with divergence of the dipolar magnetic field lines controlling the emitting plasma flow. We believe that the most of the data points lying at or close to the LBL represent the profile components (core or conal) of almost orthogonal rotators. The points well above this line in Figure \ref{figure.2} represent either geometrical broadening (due to significant non-orthogonality) or larger parts of the entire pulse profile (in most cases full widths), or both. However, points above the LBL in Figure \ref{figure.3} (presenting exclusively the widths of the conal profile components), represent only the geometrical broadening due to small inclination angles.

\begin{figure}
\begin{center}
\includegraphics{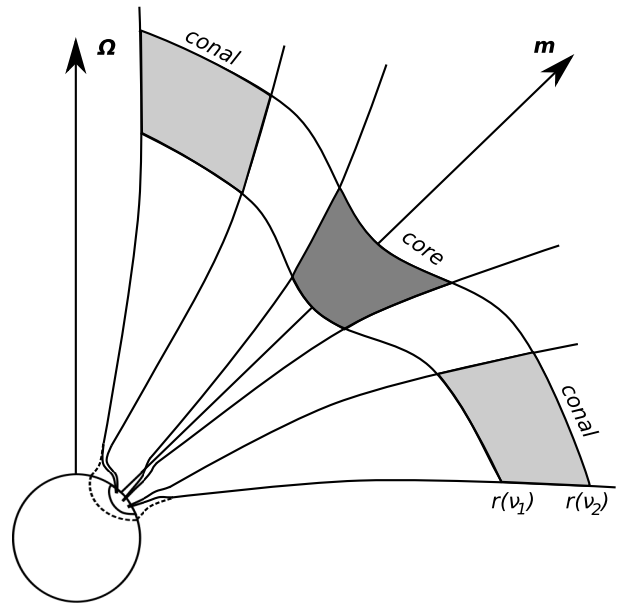}
\caption{Schematic illustration of polar cap and radio emission regions with different altitudes for core and conal components. It is assumed that the emission altitudes  $r(\nu_1) < r(\nu_2)$ for $\nu_2 > \nu_1$. The magnetic field is purely dipolar in the radio emission region, while near the polar cap the surface field is dominated by strong crust anchored magnetic anomaly. The streams of electron-positron plasma created by polar cap sparks are driven by non-dipolar magnetic field near the surface, which smoothly adopt purely dipolar form far from the surface. The geometrical dipolar polar cap is marked by dashed line, while the actual non-dipolar polar cap is marked by solid line.
\label{figure.5}}
\end{center}
\end{figure}

To explain the nature of the intrinsic beam structures that can manifest themselves in the form of LBL one has to consider a relationship between instantaneous and average pulsar emission. In morphological terms it translates into a connection between subpulses in single pulses and profile components in the average profile. In this paper we adopt the following picture. The subpulses are produced in the radio emission region within columns of plasma initiated at the polar cap surface by sparks discharging the inner accelerating potential drop. Although the subpulse emission results from incoherent addition of a large number of elementary shots of coherent radiation (e.g. \citet{gs90}; \citet{gil04} and references therein), its width is determined by the angular extent of the bundle of dipolar field lines controlling the flow of the subpulse-associated plasma flows. The sparks populate the surface of the polar cap as densely as possible, with each spark keeping an approximately the same distance from its neighbours. This is because the sparking discharge of the polar gap occurs in every place where the accelerating potential drop exceeds the threshold for the cascading electron-positron pair production. It is well known that the magnetic field in the radio emission region must be purely dipolar (at altitudes $r_{em}$ exceeding 10 neutron star radii). However, the surface magnetic field at the polar cap should be very strong and curved to ensure efficient pair production. Therefore, the gap magnetic field must be highly non-dipolar and the actual polar cap surface area should be compressed to a small fraction of the canonical polar cap (see Section \ref{section.beam.structures} for details and Figure \ref{figure.5} for illustration). The cross-section of sparks populating the actual polar cap should also be very small but the plasma columns created by these sparks will progressively expand in lateral dimensions, while they approach and enter the radio emission region. Then it is convenient to make all necessary estimates at the radio emission altitudes. We assume that the observed width of the profile component is determined by the angular extent of the bundle of dipolar magnetic fields encompassing (at any instant) the spark-associated plasma flow corresponding to radio emission related to that component. For simplicity of calculations we adopt that the impact angle $\beta$ is small (close to $0^{\circ}$) and the inclination angle $\alpha$ is large (close to $90^{\circ}$). The first assumption ignores the component broadening due to the conal structure of the average beam, while the second assumption ignores the component broadening due to small inclination angles (approximately like $1/\sin\alpha$). We will discuss the influence of both these factors later on.

Yet another assumption we make is that the core and the conal emission is generated within the plasma columns at approximately the same altitude of about 40 -- 50 stellar radii in a typical pulsar. We believe that this assumption is supported by the data presented and discussed in this paper, although we cannot judge how closely it is realised in nature (for example, see the aberration/retardation (A/R) constraints discussed at the end of Section \ref{section.explanation.r90}). Anyway, in our model illustrated in Figures \ref{figure.4} and \ref{figure.5} all beams of radiation (both inner/core and outer/conal) originate in the region of dipolar magnetic field at altitudes $r_{em}=r_6 R$, where $r_6 \gg 10$ (see Section \ref{section.constraints}). Although in our opinion there is no significant difference between core and conal emission, in the following we consider these components separately, just for clarity of presentation (in Section \ref{section.constraints} we discuss a possibility that the core emission originates at slightly lower altitudes than the conal one). In Figure \ref{figure.4} we present schematically the cross-section of the open dipolar field lines at altitudes of the radio emitting region. The parameter $s$ labels different field lines from $s=0$ for the dipolar axis to $s=1$ for the last open field lines. Three Gaussian profiles represent three subbeams of radio emission associated with three different plasma flows. The regions marked in grey correspond schematically to the mean pulsar emission. The conal emission forms a ring due to slow motion of the instantaneous beams around the magnetic axis \citep{gks93,dr99,gupta04}. This motion, which is best explained by the sparking gap model (Ruderman \& Sutherland (1975); \citet{rs75} hereafter; \citet{gil03}), manifests itself observationally by the subpulse drifting phenomenon. The core mean emission is more localised and forms a pencil beam marked by dark grey.

In the following we assume that the coherent pulsar emission is relativistically beamed along dipolar magnetic field lines. The opening angle of emission between the magnetic axis and the tangent to a dipolar field line labelled by the parameter $s$ can be written in the form
\begin{equation}
     \rho = 1.24^{\circ} s r_6^{0.5} P^{-0.5},
\label{eq.1}
\end{equation}
(e.g. \citet{gk93}). This expression is general and holds for both core and conal emission. For the bundle of the field lines encompassing the plasma flow we have the angular extent (intrinsic width)
\begin{equation}
     \Delta\rho = 1.24^{\circ} \Delta s r_6^{0.5} P^{-0.5},
\label{eq.2}
\end{equation}
where $\Delta s$ is the range of labelling parameters $s$ covered by the base of the bundle at the polar cap. It is important to keep in mind the observed width $\Delta W \ge \Delta \rho$, where the equality sign corresponds to the cases of nearly orthogonal rotators ($\alpha\approx90^{\circ}$) and nearly central cuts of the line-of-sight ($\beta\approx0^{\circ}$).

\subsubsection{Core component}\label{subsection.core}
For the core component (marked as the inner Gaussian in Figure \ref{figure.4}) the labelling parameter $s\approx0$ near the maximum and a value of $s\ll1$ correspond to 50 per cent intensity level. Therefore, according to Equation (\ref{eq.1}) $W_{50}=2\rho=2.48^{\circ} s r_6^{0.5} P^{-0.5}\approx2.5^{\circ} P^{-0.5}$. The right-hand side of this equation describes the LBL visible in Figure \ref{figure.2}. We get immediately the condition $s r_6^{0.5}\approx1$. Let us assume that the emission altitude for core components is about the same as for the conal ones, thus $r_6\approx$ 20 -- 80 for the range of periods from 0.1 to 2 s (see Figure 2 in \citet{kg98}). This implies the range of values of the parameter $s$ between 0.1 and 0.2, meaning that the core emission beam covers about 30 per cent (more exactly between 20 -- 40) of the radius of the bundle of open magnetic field lines.

\subsubsection{Conal components}\label{subsection.conal}
For the conal components we can use Equation (\ref{eq.2}) and write $W_{50}\approx\Delta\rho=\rho_2-\rho_1=1.24^{\circ} (s_2-s_1) r_6^{0.5} P^{-0.5} = 1.24^{\circ} \Delta s r_6^{0.5} P^{-0.5} = 2.5^{\circ} P^{-0.5}$, where again the right-hand side corresponds to the observed LBL. Thus, we get the condition $\Delta s = 2 r_6^{-0.5}$, which for $r_6\approx$ 20 -- 80 (see \citet{kg98} and Section \ref{section.constraints}) gives $\Delta s \approx 0.3$ (within a range approximately 0.2 -- 0.4). The cross-section of the conal beam covers about 30 per cent of the radius of the bundle of open magnetic field lines.\\

The observed LBL will correspond to the narrowest features that can be distinguished in the average pulsar beam. They will represent core or conal components in nearly orthogonal rotators. It follows from the above considerations that the LBL will have a form $W_{50}\approx 2.5^{\circ} P^{-0.5}$ provided that $\Delta s \sim 0.3$ and $r_6 \sim 50$. This new interpretation of the LBL is consistent with the analysis of the shape of pulsar beams published by Mitra \& Deshpande (1999; hereafter MD99). They explored the picture found by \citet{r93} and confirmed by \citet{gks93} and \citet{kramer94}, according to which the mean pulsar emission is organized into multiple cones with angular radii $\rho(W_{50}, \alpha, \beta)\approx4.3^{\circ}P^{-0.5}$ and $5.7^{\circ}P^{-0.5}$ for the inner and the outer cone, respectively. \citet{md99} revisited this problem and found that the pulsar emission beams follow a nested cone structure with at least three distinct cones, although only one or more of them may be active in a given pulsar (in our Figure \ref{figure.4} only one conal ring is marked for simplicity). Therefore, the angular distance between the subsequent cones ($1.4^{\circ} P^{-0.5}$) may be smaller than the cone width itself ($\sim 2.4^{\circ} P^{-0.5}$). Most importantly, \citet{md99} found that each emission cone is illuminated in the form of an annular ring of width about 20 per cent of the cone radius. Although \citet{md99} did not realized it, this implied an existence of the LBL in the form $W_{50}\sim 2.4^{\circ} P^{-0.5}$. Indeed, according to Equation (\ref{eq.2}) if $\Delta s\sim0.2$ then $W_{50}\sim0.25^{\circ} r_6^{0.5} P^{-0.5}$, which gives $W_{50}\sim2.4^{\circ} P^{-0.5}$ for $r_6\sim100$. For simplicity we used a model with one conal ring. If the actual number of conal rings is larger (2 or 3) then in such pulsars the actual value $\Delta s$ will be closer to the lower limit of the range 0.2 -- 0.4 that we found. This is fully consistent with conclusion of \citet{md99} who argued that $\Delta s \sim 0.2$.

Now we discuss our two simplifying assumptions mentioned above, namely small impact angle $\beta$ and large inclination angle $\alpha$. \citet{gks93} were the first to propose that the conal structure of the average pulsar radio beam results from the rotation of the outer instant beams around the magnetic axis. Nowadays this model is commonly accepted and known as the pulsar carousel model, thanks to extended works of Joanna Rankin and Avish Deshpande (e.g. \citet{dr99}). If the impact angle $\beta$ is small, then measuring the pulse-width of the conal component we estimate approximately the angular extent of the instant emission beam at the position of this component. However, for larger impact angles we begin to scan the cone of the emission, which in general will result in larger width that can lie above the LBL. This is illustrated by the dashed l-o-s in Figure \ref{figure.4}.

Another factor broadening the apparent pulse-width of conal beams is small inclination angle $\alpha$. \citet{r90} argued that this introduces $1/ \sin\alpha$ broadening factor in core component widths. We believe that the situation is the same in conal components, provided the impact angle $\beta$ is small\footnote{It is worth emphasizing here that in the original method of \citet{r90} developed for core components the actual value of $\beta$ can be neglected, while in the case of conal components $\beta$ has to be small. This is related to the fact that the conal beam has Gaussian intersection only when $\beta$ is small. In practice this method cannot be applied for conal single profiles or conal double profiles with a shallow saddle between the components. However, it should work fine for conal components in conal triple, quadruple and multiple, as well as deep saddle conal double profiles (see \citet{r93} for profile classification scheme).}. Therefore, we propose that \citet{r90} method of estimating the value of the inclination angle $\alpha \approx \arcsin (2.5^{\circ} P^{-0.5} / W_{50})$ is valid for both core and most of conal components (for which $W_{50}$ can be measured). We believe that most of the points in Figure \ref{figure.3} illustrate usefulness of \citet{r90} method for the widths of conal components. In two cases (6 and 7) the inferred value of $\alpha=30^{\circ}$ was confirmed by independent estimates \citep{ew01}. In other cases we just did not have an access to such estimates.

Analysing the available pulse-width data we claim that the LBL apparent in figures presenting distribution of $W_{50}$ versus $P$ results from the fact that the narrowest stable structures in the pulsar beam extend over about 1/3 of its angular radius. In Section \ref{section.beam.structures} we argue that this a natural consequence of physical processes at the pulsar polar cap. According to Equation (\ref{eq.1}), the resulting pulse-width $W_{50} =1.24^{\circ} \Delta s r_6^{0.5} P^{-0.5} / (\sin\alpha)$, which equals to the observationally derived expression $2.5^{\circ} P^{-0.5}/ (\sin\alpha)$ if $\Delta s r_6^{0.5}\approx 2$ (taking into account the measurement uncertainties $(2.4\pm1)^{\circ} P^{-0.5}$ this condition should have the form $\Delta s r_6^{0.5} \approx 2\pm 0.8$). With $\Delta s\approx 1/3$ we obtain $r_6 \approx 50$ in a typical pulsar. This should be almost independent of the pulsar period, though both $r_6$ and $\Delta s$ can be slightly period dependent. For example, in semi-empirical model \citet{kg97,kg98} argued that $r_6\propto P^{0.3}$ thus $r_6^{0.5} \propto P^{0.15}$. This would require that $\Delta s \propto P^{-0.15}$, implying that longer period pulsars can have more complex organization of their emission beams than the shorter period pulsars, despite wider beams of the latter. Although it seems that in normal (non-millisecond) pulsars this is really observed, we will not attempt a full discussion of the beam structure and pulse profile morphology. This problem is beyond the scope of our paper and we postpone it to a future Paper IV.

\subsection{Formation of pulsar beam structures}\label{section.beam.structures}
In this section we shortly discuss physical processes taking place near the polar cap surface and argue that they naturally lead to such an organization of pulsar beams that is manifested observationally by the LBL in the form $W\approx 2.5^{\circ}P^{-0.5}$. Our theoretical frame is based on the Partially Screened Gap (PSG) model of the inner acceleration region above the polar cap \citep{gil03,gil08}. This is, to the best of our knowledge, the only model of the inner accelerator that can lead to generation of the coherent radio emission in pulsars. The PSG model is a modification of Vacuum Gap (VG) model introduced by \citet{rs75}. This prototype VG model was very innovative and inspirational but had a number of shortcomings that precluded it from being a credible model of inner pulsar accelerator. Firstly, the potential drop exceeding $10^{12}$ V in pure vacuum was too high, which resulted in too fast a subpulse drift as compared with observations and too hot a surface of the polar cap due to back flow heating (see \citet{gil08} for more details). Secondly, although \citet{rs75} realised that radius of the curvature of magnetic field lines has to be about $10^6$ cm to drive a discharge of VG, they never used explicitly a non-dipolar surface magnetic field in their paper. It also appeared that the cohesive energy in typical magnetic field of about $10^{12}$ G was too low to keep charged particles in the crust and form VG above the polar cap. It is then obvious that the creation of the inner pulsar accelerator necessary for the radio emission mechanism requires strong and curved magnetic field at the polar cap, deviating largely from the global dipolar component. Only recently \citet{medin06,medin07} showed that vacuum gap can form above neutron star polar cap if the surface magnetic field is about $10^{14}$ G. Earlier \citet{gil03} proposed the PSG model in which they postulated that surface magnetic field in all radio pulsars is in a form of strong ($B_s\sim10^{14}$ G) and curved ($R_{c}\leq 10^6$ cm) magnetic anomalies, with approximately the same values of $B_s$ and $R_{c}$ in all radio pulsars\footnote{A possible mechanism of generation of such a strong, non-dipolar surface magnetic field structures is the Hall instability occurring in the neutron star crust \citep{grg03,pg10}}. This small scale non-dipolar surface magnetic field connects smoothly with the large scale dipolar field at altitudes of a few neutron star radii. This is illustrated in Figure \ref{figure.4} (for details see \citet{gmm02}). Under such assumed conditions the actual polar cap defined as the locus of the open magnetic field lines is much smaller as compared with the canonical dipolar \citet{goldreich69} picture. Due to conservation of the magnetic flux carrying by the open magnetic field lines, the classical radius of the polar cap $r_{GJ}=1.4 \times 10^4 P^{-0.5}$ cm decreases about 10-fold. Indeed, the ratio of the surface area $A_{pc}/A_{GJ}=B_d/B_s=(10^{12} \dot{P}_{-15}^{0.5} P^{0.5})/10^{14} \approx 10^{-2} (P \dot{P}_{-15})^{0.5}$. Therefore, the radius of actual polar cap in typical pulsar is only about 15 meters, as compared to 150 meters in the canonical Goldreich-Julian dipolar case. Of course, the cross-section of spark avalanches driven by curved surface magnetic field lines must be small as well, so large number of them could operate at the same time on this tiny polar cap. Each small spark feeds a column of plasma above it, which expands in lateral dimension as the magnetic field becomes more and more dipolar towards the radio emission region. Each column correspond to the observed subpulse when intersected by the observer's line of sight. Slow drift of spark discharges in the gap region makes on the average formation of the multiconal structure of the mean pulsar beam. These conal rings have been postulated by Rankin (1990, 1993) and then analysed by \citet{md99}, who found their angular widths being about 20 per cent of the angular radius of the pulsar beam. All the above aspects are schematically illustrated in Figure \ref{figure.5}. Note that the high accelerating potential drop exist only along open magnetic field lines originating at the actual polar cap (marked by solid line circle). In reality this cap is much smaller than the geometrical \citet{goldreich69} polar cap (marked by dashed line circle), but the actual scale could not be kept in this figure.

Let us consider how many discharging sparks can operate on such small polar cap within the PSG model. For typical values $B_s\sim 10^{14}$ G and $T_s\sim10^6$ K (see Table 1 in \citet{gil08}), the height $h$ of the PSG is about 30 meters \citep{szary11}. It is reasonable to adopt that the radius of curvature of the surface magnetic field $R_{c}$ is about $10^6$ cm. We can now estimate the lateral spread of the sparking avalanche developing via magnetic pair production by curvature $\gamma$ photons emitted by electrons and positrons travelling relativistically along curved magnetic field lines. During the first generation the avalanche will grow in the direction perpendicular to field lines by about $h^2/R_{c}$, which is about 9 cm for the adopted values of $h$ and $R_{c}$. It takes about 40 generations to reach the corotational limited value of the Goldreich-Julian (1969) charge density\footnote{This estimate of a number of generations necessary to regain the corotational charge density within the gap was first made by \citet{rs75} and it is still valid even for PSG accelerator. It is worth noting that \citet{rs75} also assumed curved non-dipolar magnetic field in the gap with the radius of curvature of about $10^6$ cm.}. After this time the charge density is fully replenished, the gap potential drop disappears and the avalanche stops growing. Therefore, the cross-section of fully developed spark is about 3.6 meters on the polar cap with a radius of about 15 meters. Thus, in a typical pulsar about 20 sparks can operate on the surface of the actual polar cap. This leads to $\Delta s$ about 0.25, meaning that size of each spark covers about 1/4 of the polar cap radius (which is consistent with independent estimate of \citet{cheng77}, who found the spark fractional area between 0.01 and 0.1 of the polar cap surface ($[(1/3)/2]^2\sim0.03$)). This theoretically estimated spark dimension is consistent with the angular dimensions of the core and conal components in the overall pulsar beam. Indeed, for the normalised altitudes $r_6$ of the core and the conal emission being about 60 ($r_{em}\sim600$ km) and $\Delta s \approx 0.25$, Equation (\ref{eq.2}) gives $W_{50}\approx\Delta \rho \approx 2.4^{\circ} P^{-0.5}$, consistent with the observed LBL, which reflects the angular size of the narrowest stable structures in the pulsar beam.

In summary, the ultra strong accelerating potential drop in PSG is discharged by the number of localised sparks (avalanches), each covering a surface area of the polar cap with a characteristic dimension of about $\Delta s = 0.25$ of the polar cap radius. These sparks deliver streams of the electron-positron plasma to the radio emission region. As argued in Sections \ref{subsection.core} and \ref{subsection.conal} the emission beams associated with these streams cover just about 25 per cent of the radius of overall emission beam. This is consistent with the already mentioned analysis of \citet{md99}, who demonstrated that each conal ring illuminates about 20 per cent of the pulsar beam radius. Therefore, the theoretical estimate of angular size of the spark-associated plasma flows and the observationally deduced angular extent of stationary structures in the mean pulsar beam are consistent with each other. These structures will manifest itself as the LBL in the scatter plot of the component width versus the pulsar period, provided that the core and the conal components are emitted at about the same altitudes of the order of 500 km. One should however keep in mind that according to the observed aberration/retardation (A/R) effects the core component is likely to originate about 100 km below the conal components \citep{bcw,mitra04,krzeszowski09}. Indeed, a typical A/R shift between core and conal components is of the order of 10 millisecond, with the core components being late with the respect to the profile midpoint. A possible reason of these effects is shortly discussed at the end of the next section.

\section{Constraints for the altitude of coherent radio emission} \label{section.constraints}
\textbf{
Let us begin with specifying our general assumptions, both geometrical and physical (see Section \ref{section.explanation.r90} for details). The mechanism which is responsible for the observed pulsar radio emission must be the coherent process associated with some relativistic plasma turbulence. In other words it must be some kind of collective process involving a great number of charged particles radiating in phase. The known examples of such processes in pulsar astrophysics are: cyclotron maser \citep{kmm91}, two-stream instabilities \citep{u87,am98}, collapsing solitons \citep{w98}, charged relativistic solitons \citep{mgp00,gil04} and linear acceleration maser \citep{m78}.
}

It is obvious that such a process should take place in the magnetospheric pair plasma that moves relativistically along the pulsar dipolar magnetic field lines. A number of necessary conditions must be fulfilled in order to generate an observed pulsar radio emission, irrespective of the actual coherence mechanism. It should be expected that favourable conditions will occur above some minimum altitude, which we intend to determine below. We will concentrate at high frequency radiation $\nu\sim1$ GHz, because lower frequencies are most likely emitted at higher altitudes.

Let us assume that the radiowaves with the characteristic wavelength $\lambda_{em}$ are generated at some altitude $r_{em}$ in a region with characteristic longitudinal (along magnetic field lines)\footnote{\textbf{The perpendicular dimension of the radiation region is not important for our considerations, although it could be similar to the longitudinal one or even larger. The pulsar radiation is emitted tangentially to the diverging magnetic field lines and only the adjacent sources can contribute to the incoherent addition of shots of the coherent emission.}} scale $l_{em}$ (not to be confused with the longitudinal size $\Delta$ of the elementary emitter), where both $\lambda_{em}$ and $l_{em}$ are measured in the laboratory (observer's) frame of reference. The plasma in which the radio emission is generated is moving with the characteristic Lorentz factor of about  $\gamma_{p}=\left(2-5\right) \times 10^{2}$ with respect to the laboratory frame. 

%Despite the details of radiation mechanism in the co-moving frame of reference (that moves with the plasma Lorentz factor) the following general condition has to be satisfied

Despite details of radiation mechanism the characteristic size of the emission region $l_{em}$ must be much larger than the wavelength of the emitted radiation (which in term must be larger than the longitudinal size of elementary emitters), i.e. $\Delta < \lambda_{em}\ll l_{em}$. Otherwise, one could not consider any collective plasma processes leading to generation of the coherent radiation. In other words this condition reflects applicability of the geometrical optics approximation (eikonal approximation; e.g. \citet{rl79}, p.73), which should be valid for any radiation mechanism, including the one leading to the observed pulsar radio emission\footnote{\textbf{Of course, one can think of some man-made antenna mechanism with the size of antenna smaller than the emitted wavelength (typically half of it) but they do not occur in nature.}}. Of course this condition must
also be satisfied in the co-moving frame of reference 

\begin{equation}
\lambda_{em}^{\prime }\ll l_{em}^{\prime }.  \label{eq.3}
\end{equation}
According to the Lorentz transformation $\lambda_{em}^{\prime }=\lambda_{em}\gamma_{p}$ (relativistic Doppler effect), while due to Lorentz contraction $l_{em}^{\prime }= l_{em}/\gamma_{p}$. Thus, Equation (\ref{eq.3}) can be rewritten in the observer's frame in the form
\begin{equation}
\lambda_{em}\ll\frac{1}{\gamma_{p}^{2}}l_{em},  \label{eq.4}
\end{equation}
which will be useful in deriving the lower limits for the altitude of the pulsar radio emission. For gigahertz frequencies ($\lambda_{em}\sim10 \text{cm}$) and assuming reasonable $\gamma_{p}^{2}\sim10^{5}$ one obtains 

\begin{equation}
l_{em}\gg\gamma_{p}^{2}\lambda_{em}\sim10^{7} \text{ cm}. \label{eq.5}
\end{equation}
Since the altitude of the emission region $r_{em} > l_{em}$, this means that the observed pulsar radio emission cannot originate below $10^7$ cm, irrespective of the actual mechanism of generation of the coherent pulsar radiation. This excludes any radio emission originating at or near the polar cap, as proposed by \citet{r90} for the core beam. The core emission region must be located at altitudes largely exceeding $10^7$ cm, perhaps close to the conal emission region at altitudes \begin{equation}
r_{em}^{conal}\sim 50 R \nu_{\text{GHz}}^{-0.25} P^{0.33}, \label{eq.6}
\end{equation}
which is about 50 stellar radii $R=10^6$ cm for a typical pulsar \citep{kg97,kg98}.

The above conclusion is general, and it is valid for any possible mechanism of generation of the coherent pulsar radio emission. Let us now assume that this emission is generated by the coherent curvature radiation, which seems to be most relevant mechanism \citep{gil04,mitra09}. 

%\textbf{First, let us note that the condition for the coherent radio emission $\Delta < \lambda_{em}$, where $\Delta$ is a size of elementary emitter, does not contradict the general eikonal condition expressed by Equation (\ref{eq.3}).}

Thus $\lambda_{em}$ is a function of the radius of curvature of dipolar magnetic field lines in the emission region, which can 
be expressed in the form

\begin{equation}
R_{c}=\frac{R}{3}\frac{\sin \theta }{\sin ^{2}\theta_{\ast }}\frac{\left(1+3\cos ^{2}\theta \right) ^{1.5}}{1+\cos ^{2}\theta },  \label{eq.7}
\end{equation}
$\theta_{\ast}$ and $\theta $ are the polar coordinates of the same field line at the stellar surface and at the radio emission altitude $r_{em}$, respectively. For the open field lines $\theta <\theta_{m}$, where $\theta_{m}=1.\,45\times 10^{-2} P^{-0.5}$ is the opening angle of the last open field line (thus it defines the opening angle of the dipolar polar cap) and $P$ is the pulsar period in seconds. 

Using the equation for the dipole field lines $\sin \theta =\left( r_{em}/R\right) ^{0.5}\sin \theta _{\ast }$ one can rewrite Equation (\ref{eq.7}) in the form

\begin{equation}
R_{c}=\allowbreak 9.\,\allowbreak 21\times 10^{7}\dfrac{r_{6}^{0.5}}{s}P^{0.5},  \label{eq.8}
\end{equation}
where $r_{6}=r_{em}/R$ and $s=\theta _{\ast}/\theta_{m}$ (labelling parameter in Equation (\ref{eq.1}), see also Figure \ref{figure.4}). The curvature radiation reaches the maximum at frequency $\nu_c = 0.44 (c/2\pi)\gamma^3_p/R_c$ (\citet{rl79}, p. 179), thus
\begin{equation}
\nu _{c}=2.1\times 10^{9}\allowbreak \gamma _{p}^{3}/R_{c}. \label{eq.9}
\end{equation}
Now for $\lambda_{em}=c/\nu_c$ we can rewrite the general Equation (\ref{eq.4}) in the form

$$ l_{em}\gg\frac{14.3}{\gamma_{p}}R_{c}, $$
or using Equation (\ref{eq.8})
$$l_{em}\gg1.\,\allowbreak 317\times 10^{9} \dfrac{r_{6}^{0.5}}{s\gamma_{p}}P^{0.5}. $$
Since $l_{em} \lesssim r_{em}=r_{6} R$ cm then we obtain condition for the radio emission altitude expressed in units of the stellar radius $R=10^6$ cm in the form
\begin{equation}
r_{6}\gg1.\,\allowbreak 7\times 10^{6} \dfrac{P}{s^{2}\gamma _{p}^{2}}.
\label{eq.10}
\end{equation}
Since $s<1$ and $\gamma^2_p\sim10^{5}$ we have $r_{6}\gg17 P$, which means that the coherent curvature radiation can be generated at the altitudes well exceeding $10^7$ cm, no matter whether core or conal emission is considered.

We can discuss the sensitivity of above estimated constraint on the adopted assumptions: (i) dipolar magnetic field, (ii) $s<1$ and (iii) $\gamma_p=(2-5) \times10^2$. The first assumption is well justified for normal pulsars. The second assumption means restriction to the open dipolar magnetic field lines. We can only ask how large the Lorentz factor $\gamma_p$ must be to bring the emission region to the vicinity of the polar cap $r_6\sim2$. It is easy to see that $\gamma_p$ would have to be larger than 1000, which does not seem possible. 

It is more convenient to define $\gamma _{p}$ as a function of a frequency $\nu_{\text{GHz}}=\nu _{c}/10^{9}$. Using the expression $\gamma _{p}=352.65\left(\nu_{\text{GHz}} P^{0.5} r_{6}^{0.5} / s\right)^{\frac{1}{3}}$ we can rewrite Equation (\ref{eq.10}) in the form
\begin{equation}
r_{6}\gg7.1  s^{-1} P^{0.5} \nu_{\text{GHz}}^{-0.5}, \label{eq.11}
\end{equation}
which describes the magnetospheric region physically forbidden for generation the coherent curvature radio emission. Even if $s\sim1$ for the edge of the core emission (as assumed by \citet{r90} this radiation cannot originate below about $10^7$ cm). Thus, we conclude again that the observed pulsar radiation (no matter core or conal) can be generated at altitudes well exceeding $10^7$ cm. This conclusion is now based on the assumption that the pulsar radio emission mechanism is the coherent curvature radiation. Although, this matter is still debated, our conclusion is consistent with Equation (\ref{eq.5}), which is independent of the actual mechanism of coherence of pulsar radio emission.

Let us note that the above conditions (\ref{eq.10}) and (\ref{eq.11}) are well consistent with the radio emission altitudes for conal components ($s\sim1$) expressed by Equation (\ref{eq.6}), which is about 500 km for a typical pulsar. For the curvature radiation emitted at the altitude $r_6$ (frequency $\nu(r_6)$) by charged sources moving relativistically with the Lorentz factor $\gamma_p$ along dipolar field lines with radius of curvature $R_c$, the parameter $s$ has a very important meaning. Indeed, between the outer (conal $s\sim1$) and the inner (core $s\sim0.1$) magnetic field lines the radius of curvature (Equation (\ref{eq.8})) increases by a factor of several. Thus, the emission altitude $r_6$ should decrease by a factor of about 3 or more to compensate the change caused by varying $s$. Therefore, the core (central) emission should originate at slightly lower altitudes than the conal (outer) emission, however not closer to the polar cap than the conditions (\ref{eq.10}) or (\ref{eq.11}) do allow. It is then natural that the core emission originates about 100 km lower than the conal emission. This seems to be confirmed by the observed A/R effects (e.g. \citet{mitra04,mrg07,krzeszowski09}). A schematic illustration is presented in Figure \ref{figure.5}, where the core emission region is located at slightly lower altitudes than that of the conal emission. It is assumed that $\nu_1>\nu_2$ and radius-to-frequency mapping holds for conal components in agreement with Equation (\ref{eq.6}).

\section{Conclusions}\label{section.conclusions}
Our main conclusions are the following:
\begin{enumerate}
      \item [1.] Lower Boundary Line $W_{50}\approx2.5^{\circ} P^{-0.5}$ or rather Lower Boundary Belt $W_{50}=(2.4^{\circ}\pm1^{\circ})P^{-0.5}$ exist in scatter plot of 50 per cent pulse-width versus pulsar period for both core and conal profile components. They reflect the narrowest stable structures distinguishable within the average pulsar beam, whose angular extent is determined by physics of the acceleration region above the polar cap. These structures manifest themselves by the core and the conal components in mean pulsar profiles.
      \item [2.] The simple method of estimating the inclination angle $\alpha=\arcsin(2.5^{\circ} P^{-0.5}/W_{50})$ from the half-power pulse-width $W_{50}$ can be applied to both core and conal profile components.
      \item [3.] Both the core and the conal pulsar emission originate far from the stellar surface, presumably at altitudes of about 500 km in typical pulsars. However, it is possible that the core emission is generated about 100 km closer to the polar cap than the conal emission, which is manifested by the observed by the A/R effects.

\end{enumerate}

\section*{Acknowledgments}
This work is partially supported by the Polish Research Grant NN 203 3919 34. GM was partially supported by the Georgian NSF grant ST08/4-442. We thank dr. W. Lewandowski for critical reading of the manuscript and very helpful comments.

\bsp

\newpage

\appendix
\section{Conal component widths}\label{appendix.spec}
All pulse widths measurements used in Figure \ref{figure.3} are presented in Table \ref{table.1} and described for different blocks of data marked in the figure's key.

\subsection{Parkes Multibeam Pulsar Surveys}
We reviewed papers presenting the results of Parkes Multibeam Pulsar Surveys (hereafter PMPS I -- VI. We identified 70 PMPS profiles of normal pulsars (excluding millisecond ones), in which we were able to measure/estimate the half-power pulse-width $W_{50}$ of at least one conal component. These measurements are listed in Table \ref{table.1} and presented in Figure \ref{figure.3} in the form of the scatter plot. Different colours correspond to different PMPS surveys, as marked in the key.
\subsection{Partial Cone Profiles}
\citet{lm88} first identified a group of 50 pulsars, in which a part of their conal profiles were apparently missing. Recently \citet{mr11} reinvestigated this population of pulsars using new sensitive polarimetric observations (Arecibo and GMRT) and argued that the missing conal components were just weak and appeared in occasional bursts. These pulsars are especially useful for our purposes. Indeed, in most of them at least one conal component is sufficiently prominent to make an estimate of its half-power width possible. In some cases the single conal component dominates the full profile width measurement (see Figure 3 in \citet{mg11}). In 39 profiles published by \citet{mr11} we found 16 cases in which we succeeded to obtain a credible estimate of half-power width of strong conal component. These measurements are listed in Table \ref{table.1} and presented in Figure \ref{figure.3} as dark blue dots. Most points lie near the LBL suggesting the inclination angle $\alpha > 45^{\circ}$ (see the dotted line representing the method of \citet{r90} and Figure 5 in \citet{mg11} with exact calculations). This seems to be consistent with the estimates of $\alpha$ given in Table A3 of \citet{mr11}. There are two pulsars in this table (B1322+83 and B2224+65) with small values of $\alpha= 14^{\circ}$ and $27^{\circ}$, respectively. These two pulsars lie in our Figure \ref{figure.3} at the largest distance from the LBL (but below the dotted line corresponding to $\alpha= 15^{\circ}$). When we applied the \citet{r90} method $\alpha=\arcsin(W_{50}/(2.37^{\circ} P^{-0.5}))$ we obtained $11^{\circ}$ and $17^{\circ}$, respectively\footnote{As shown in \cite{mg11} (see Figure 6 there), the accuracy of \citet{r90} method decreases for smaller values of the inclination angles.}. This suggests that the \citet{r90} method can be used for the conal profile components with the same accuracy as for the core ones. Below we show more of the special cases supporting this suggestion.
\subsection{Special Cases}
We identified 10 special cases marked in magenta and labelled by capital letters or numbers. They have a special meaning for our reasoning, which we describe in some detail below in subsections 2.3.1 -- 2.3.8. In general, we estimated $50 \%$ pulse-widths of different components in complex profile pulsars. Whenever possible we intended to check whether there are some significant differences in the pulse-width between the core and the conal components. We introduced the following scheme to denote different components: $\bigtriangleup$ - leading outer conal, $\circ$ - leading inner conal, $+$ - core, $\Box$ - trailing inner conal and $\Diamond$ - trailing outer conal.
\subsubsection{Conal component in DP-IP cases}
For all 31 DP-IP cases presented in Figure \ref{figure.2} as the red dots their pulse-width measurements corresponded to the core components. In most of these cases the conal components were either non-existent or impossible to distinguish. However, in the profiles of two interpulsars: B1055-52 and B1822-09 (marked in Figure \ref{figure.3} by letters A and B, respectively) we were able to distinguish the conal component (see Figures A3 and A6 in \citet{mr11}) and estimate its half-power pulse-width. These measurements are presented as magenta stars associated with red-coloured DP-IP symbols. It is important to note that the conal width estimates fall within the error bars of the core width measurements in each of the two presented cases. This suggests that in complex profile pulsars core and conal components have pulse-widths very close to each other (below we present more evidence to this otherwise intuitive suggestion). Thus, it is quite possible that the conal components have the same lower boundary as the one found for the core components by \citet{r90} and confirmed in our \citet{mg11}.
\subsubsection{B1913+16}
The conal components of pulsar B1913+16 are represented in Figure \ref{figure.3} by two magenta symbols marked by the number 1. They both lie at or very close to the LBL. This short period $P=0.056$ s pulsar is an important supporting case in arguing that conal components have the same lower boundary as the core ones.
\subsubsection{B0736-40}
This is an interesting case of a new transient but stable conal component discovered recently in this pulsar by \citet{karastergiou11}. This component is represented in Figure \ref{figure.3} by the magenta symbol marked by the number 2. Its location right on the LBL suggests a large inclination angle $\alpha\sim 90^{\circ}$.
\subsubsection{B1737+13}
This is a pulsar with a complex five component profile, with three of them being suitable for making measurement of $W_{50}$. Three components of pulsar B1737+13 are represented in Figure \ref{figure.3} by the magenta symbols marked by the number 3. As one can see there is no significant difference in the widths of the core (middle) and any of the conal components. All $W_{50}$ measurements lie near the LBL, so this pulsar has large inclination angle $\alpha>75^{\circ}$ (see Figure 5 in \citet{mg11}). It is worth emphasizing that the quality of this data is very good, as the component widths have been measured by \citet{krzeszowski09}, using the Gaussian fitting method developed by \citet{kramer94}.
\subsubsection{B1237+25}
This is another five component profile pulsar, represented in Figure \ref{figure.3} by the magenta symbols marked by the number 4. Except of the trailing conal component (which is broadened by an unknown factor) all other components (including the core one) have similar widths close to the LBL value. Also in this pulsar the component widths have been measured by Krzeszowski et al. (2009; see their Figure 4) using the Gaussian fitting method. It is well known that pulsar is almost orthogonal rotator with $\alpha>75^{\circ}$.
\subsubsection{B2045-16}
The strong conal component of this complex profile pulsar is represented in Figure \ref{figure.3} by the magenta symbol marked by 5. We measured the pulse-width of one dominating conal component and $W_{50}$ lies right on the LBL. This pulsar must be therefore an almost aligned rotator, with the inclination angle $\alpha$ smaller than 75$^{\circ}$.
\subsubsection{B0301+19 and B0525+21}
These pulsars have two conal components joined by the saddle. Good quality polarimetric observations were published by \citet{ew01} and we estimated half-power widths of the conal components from their profiles. Both the measurement values (two magenta symbols marked by the numbers 6 and 7 in Figure \ref{figure.3}) lie significantly off (but above) the LBL. This suggest that the profile is broadened geometrically by the inclination angle $\alpha$ close to 30$^{\circ}$ (see caption of Figure \ref{figure.3}). Interestingly, this value is close to the independent polarimetric estimates of $\alpha$ in these pulsars (see Table 3 in \citet{ew01})
\subsubsection{J0631+1036}
This is an interesting pulsar with four symmetrical conal components (see Figure 2 in \citet{waa+10}). Two inner components are narrower than the lower boundary value, while the leading outer component lies exactly at the LBL (see the magenta symbol marked by number 8). This is consistent with the inclination angle $\alpha$ being close to $90^{\circ}$ (Figure 3 in \citet{waa+10}). However, the narrow inner components lying below the LBL are intriguing. It is worth mentioning that in \citet{mg11} we have found few such cases belonging to the category of young energetic pulsars (see Figure 3 in \citet{mg11}). Interestingly, this pulsar also belongs to this category, and that is why we omitted the narrow inner components of J0631+1036 in Figure \ref{figure.3}, while showing them in Table \ref{table.1}. Yet another case of this category is J1718-3825, which shows a distinct conal component (Figure 17 in \citet{waa+10}) with $W_{50}=6.5^{\circ}$. This is less than the LBL value $8.95^{\circ}$ for its period $P=0.075$, and we do not show it in Figure \ref{figure.3} as well.

\begin{table*}
\centering
\begin{minipage}{180mm}
\caption{50 per cent maximum intensity pulse-widths $W_{50}$ of conal components.   \label{table.1}}
\begin{tabular}{|l|c|c|c|c|c|l|c|c|c|c|}

 \cline{1-5}\cline{7-11}
\multicolumn{1}{|c|}{Name J} & \multicolumn{1}{|c|}{Name B} & \multicolumn{1}{|c|}{Period [s]}  & \multicolumn{1}{|c|}{$W_{50}$[deg]} &\multicolumn{1}{|c|}{Ref.}& \multicolumn{1}{|c|}{ }& \multicolumn{1}{|c|}{Name J} & \multicolumn{1}{|c|}{Name B} & \multicolumn{1}{|c|}{Period [s]}  & \multicolumn{1}{|c|}{$W_{50}$[deg]} & \multicolumn{1}{|c|}{Ref.} \\
 \cline{1-5}\cline{7-11}
\multicolumn{5}{|c|}{\textbf{Parkes Multibeam Pulsar Survey I}}& & \multicolumn{5}{|c|}{\textbf{Parkes Multibeam Pulsar Survey VI}} \\ 
1232-6501 &         &0.088 & 17.  & PMPS I    &  & 0733-2345 &         &1.796 & 4.0  &  PMPS VI  \\
1245-6238 &         &2.283 & 2.3  & PMPS I    &  & 0932-5327 &         &4.392 & 1.0  &  PMPS VI  \\
1307-6318 &         &4.962 & 5.   & PMPS I    &  & 1001-5939 &         &7.734 & 1.0  &  PMPS VI  \\
1345-6115 &         &1.253 & 2.9  & PMPS I    &  & 1148-5725 &         &3.560 & 1.0  &  PMPS VI  \\
1649-4349 &         &0.871 & 10.  & PMPS I    &  & 1439-5501 &         &0.029 & 10.0 &  PMPS VI  \\
1723-3659 &         &0.203 & 11.  & PMPS I    &  & 1622-4347 &         &0.458 & 4.0  &  PMPS VI  \\
\multicolumn{5}{|c|}{}                        &  & 1758-2646 &         & 0.767 & 2.7  & PMPS VI  \\
\multicolumn{5}{|c|}{\textbf{Parkes Multibeam Pulsar Survey II}}&  &  1624-4613 &         &0.871  & 5.0  & PMPS VI \\
1755-2521 &         &1.176 & 2.5  &  PMPS II  &  & 1638-3815 &         &0.698  & 6.0  & PMPS VI \\
1806-1920 &         &0.880 & 40.  &  PMPS II  &  & 1717-3953 &         & 1.086 & 19.  & PMPS VI \\
1809-1917 &         &0.063 & 20.  &  PMPS II  &  & 1728-4028 &         & 0.866 & 6.   & PMPS VI \\
1850+0026 &         &1.082 & 2.6  &  PMPS II  &  & 1750-2043 &         & 5.639 & 6.   & PMPS VI \\
\multicolumn{5}{|c|}{}                        &  & 1827-0750 &         & 0.270 & 10.  & PMPS VI \\
\multicolumn{5}{|c|}{\textbf{Parkes Multibeam Pulsar Survey III}}& &  1836-1324 &         & 0.179 & 7.0  & PMPS VI \\
1015-5719 &         &0.140 & 27.0 &  PMPS III &  & 1841-1404 &         & 1.334 & 5.0  & PMPS VI \\
1159-6409 &         &0.667 & 28.0 &  PMPS III &  & 1845-1351 &         & 2.619 & 3.0  & PMPS VI \\
1418-5945 &         &1.673 & 2.5  &  PMPS III &  &  \multicolumn{5}{|c|}{} \\
1512-5431 &         &2.047 & 6.   &  PMPS III &  &  \multicolumn{5}{|c|}{\textbf{Partial Cones}} \\
1522-5525 &         &1.390 & 2.2  &  PMPS III &  & 0141+6009 & 0138+59 & 1.223 & 2.6  & MR11  \\
1531-5610 &         &0.084 & 9.0  &  PMPS III &  & 0358+5413 & 0355+54 & 0.156 & 8.2  & MR11  \\
1532-5308 &         &0.444 & 5.5  &  PMPS III &  & 0922+0638 & 0919+06 & 0.431 & 4.6  & MR11  \\
1542-5303 &         &1.208 & 2.5  &  PMPS III &  & 1321+8323 & 1322+83 & 0.670 & 11.1 & MR11  \\
1548-4821 &         &0.146 & 7.   &  PMPS III &  & 1532+2745 & 1530+27 & 1.125 & 2.8  & MR11  \\
1654-4140 &         &1.274 & 4.   &  PMPS III &  & 1759-2205 & 1756-22 & 0.461 & 3.06 & MR11  \\
1705-3950 &         &0.319 & 5.   &  PMPS III &  & 1912+2104 & 1910+20 & 2.233 & 1.60 & MR11  \\
1707-4729 &         &0.266 & 5.   &  PMPS III &  & 1915+1009 & 1913+10 & 0.405 & 5.0  & MR11  \\
1717-4043 &         &0.398 & 6.   &  PMPS III &  & 1917+1353 & 1915+13 & 0.195 & 7.40 & MR11  \\
1725-4043 &         &1.452 & 2.5  &  PMPS III &  & 1926+1648 & 1924+16 & 0.580 & 5.0  & MR11  \\
\multicolumn{5}{|c|}{}                        &  & 1932+2220 & 1930+22 & 0.144 & 4.9  & MR11  \\
\multicolumn{5}{|c|}{\textbf{Parkes Multibeam Pulsar Survey IV}}&  &  1941-2602 & 1937-26 & 0.403 & 3.66 & MR11  \\
0729-1836 & 0727-18 &0.510 & 3.5  &  PMPS IV  &  & 2022+5154 & 2021+51 & 0.529 & 5.1  & MR11  \\
0955-5304 & 0953-52 &0.862 & 2.8  &  PMPS IV  &  & 2055+3630 & 2053+36 & 0.222 & 5.4  & MR11  \\
1133-6250 & 1131-62 &1.023 & 12.4 &  PMPS IV  &  & 2225+6535 & 2224+65 & 0.683 & 11.6 & MR11  \\
1327-6301 & 1323-627&0.196 & 12.  &  PMPS IV  &  & 2330-2005 & 2327-20 & 1.644 & 1.5  & MR11  \\
1534-5334 & 1530-53 &1.369 & 4.   &  PMPS IV  &  &  \multicolumn{5}{|c|}{} \\
1602-5100 & 1558-50 &0.864 & 3.   &  PMPS IV  &  &  \multicolumn{5}{|c|}{\textbf{Special Cases}} \\
1604-4909 & 1600-49 &0.327 & 4.8  &  PMPS IV  &  & 1057-5226 & 1055-52 &0.197  & 4.5  & MR11 (A)\\
1653-3838 & 1650-38 &0.305 & 4.5  &  PMPS IV  &  & 1825-0935 & 1822-09 &0.769  & 3.0  & MR11 (B)\\
1703-4851 &         &1.396 & 2.2  &  PMPS IV  &  & 1915+1606 & 1913+16 &0.056  & $\bigtriangleup$ 8.  & LM88 (1)\\
1727-2739 &         &1.293 & 4.3  &  PMPS IV  &  &           &         &       & $\Diamond$ 10.       & \\
1733-3716 & 1730-37 &0.338 & 6.   &  PMPS IV  &  & 0738-4042 & 0736-40 &0.375  & $\bigtriangleup$ 3.9.& LM88 (2) \\
1738-3316 &         &0.730 & 4.5  &  PMPS IV  &  & 1740+1311 & 1737+13 & 0.803 & $\bigtriangleup$ 3.0 & K09 (3)\\
1745-3040 & 1742-30 &0.368 & 4.   &  PMPS IV  &  &           &         &       & $\circ$ 2.8          & \\
1748-2444 &         &0.443 & 3.7  &  PMPS IV  &  &           &         &       & $\Diamond$ 3.1       & \\
1749-2347 &         &0.875 & 3.   &  PMPS IV  &  & 1239+2453 & 1237+25 &1.382  & $\bigtriangleup$ 1.6 & K09 (4)\\
1750-3157 & 1747-31 &0.910 & 3.   &  PMPS IV  &  &           &         &       & $\circ$ 2.0          & \\
1801-2920 & 1758-29 &1.082 & 2.2  &  PMPS IV  &  &           &         &       & $+$ 2.0              & \\
1834-0426 & 1831-04 &0.290 & 12.  &  PMPS IV  &  &           &         &       & $\Box$ 2.0           & \\
1834-1202 &         &0.610 & 15.  &  PMPS IV  &  &           &         &       & $\Diamond$ 2.8       & \\
1853-0004(A)&       &0.101 & 7.2  &  PMPS IV  &  & 2048-1616 & 2045-16 &1.962  & $\circ$ 1.8          & LM88 (5)\\
1855+0307 &         &0.845 & 4.   &  PMPS IV  &  &           &         &       & $\Box$ 1.7           & \\
1901-0312 &         &0.356 & 6.   &  PMPS IV  &  & 0528+2200 & 0525+21 &3.746  & $\circ$ 3.           & EW01 (6)\\
1901+0331 & 1859+03 &0.655 & 3.6  &  PMPS IV  &  &           &         &       & $\Box$ 2.7           & \\
1904+1011 & 1901+10 &1.857 & 3.   &  PMPS IV  &  & 0304+1932 & 0301+19 &1.388  & $\circ$ 4.0          & EW01 (7)\\
1908+0734 &         &0.212 & 5.4  &  PMPS IV  &  &           &         &       & $\Box$ 4.0           & \\
1910+0358 & 1907+03 &2.330 & 6.   &  PMPS IV  &  & 0631+1036 &         &0.288  & $\circ$ 2.4          & W10 (8)\\
1910+0728 &         &0.325 & 4.   &  PMPS IV  &  &           &         &       & $\Box$ 2.4           & \\
1926+1434 & 1924+14 &1.325 & 3.   &  PMPS IV  &  &           &         &       & $\Diamond$ 4.3       & \\
1933+1304 & 1930+13 &0.928 & 2.6  &  PMPS IV  &  &           &         &       &                      & \\
\multicolumn{5}{|c|}{}                        &  &           &         &       &                      & \\
\multicolumn{5}{|c|}{\textbf{Parkes Multibeam Pulsar Survey V}}&  &  & &       &                      & \\
1744-3922 &         &0.172 & 6.   &  PMPS V   &  &  \multicolumn{5}{|c|}{} \\  \cline{1-5}\cline{7-11}
\end{tabular}
\end{minipage}
\end{table*}

\label{lastpage}
\end{document}